\documentclass[reprint, amsmath,amssymb,aps]{revtex4-2}

\usepackage{graphicx}
\usepackage{dcolumn}
\usepackage{bm}
\usepackage{physics}
\usepackage{amsmath}
\usepackage{amssymb}

\begin{document}

\preprint{APS/123-QED}

\title{Boltzmann Distributions on a Quantum Computer via Active Cooling}

\author{Carter Ball}
\email{cball12@umd.edu}
\affiliation{Department of Physics, University of Maryland, College Park, MD 20742, USA}
\author{Thomas D. Cohen}
\email{cohen@umd.edu}
\affiliation{Department of Physics, University of Maryland, College Park, MD 20742, USA}

\date{\today}

\begin{abstract}
 Quantum computing raises the possibility of solving a variety of problems in physics that are presently intractable. A number of such problems involves the physics of systems in or near thermal equilibrium.  There are two main ways to compute thermal expectation values on a quantum computer: construct a thermal state that reproduces thermal expectation values, or sample various energy eigenstates from a Boltzmann distribution of a given temperature. In this paper we address the second approach and propose an algorithm that uses active cooling to produce the distribution. While this algorithm is quite general and applicable to a wide variety of systems, it was developed with the specific intention of simulating thermal configurations of non-Abelian gauge theories such as QCD, which would allow the study of quark-gluon plasma created in heavy-ion collisions.
\end{abstract}

\maketitle

\section{\label{sec:level1}Introduction}
Quantum computers promise to open the door to many calculations that in practice can be performed using computers that behave classically\cite{Feynman}. Over the years, significant research has gone into the development of physical quantum computers \cite{exp1,exp2}, the theory behind them \cite{QC1,QC2}, and the design of quantum algorithms \cite{alg}. There are many open questions in physics for which it is hoped that once sufficiently powerful quantum computers are developed they may provide important insights. One significant class of such problems involve non-perturbative non-equilibrium dynamics \cite{Snowmass}.

This paper describes an algorithm that uses the method of active cooling to create a Boltzmann distribution. Furthermore, we demonstrate that this algorithm can predict the energy of the system of interest and so, given a mapping between energy and temperature, the algorithm can be used to target specifically chosen temperature ranges. The proposed algorithm can thus be used to sample energy eigenstates from a Boltzmann distribution of a chosen temperature range.

While the algorithm we develop in this paper is rather general, our motivation is at least in part to develop an approach suitable for applications to thermal properties of non-Abelian gauge theories, such as quantum chromodynamics (QCD), as would be needed for studies of heavy-ion reactions. As with most quantum computing approaches to quantum field theories, the formalism considered here is in the context of Hamiltonian dynamics, rather than a Lagrangian/action based formalism commonly used in Euclidean space-time. For the purposes of the discussion here, it does not matter whether one wishes to think of the dynamics as being realized on a quantum simulator where interactions are constructed to mimic the dynamics of interest or via a gate-based general quantum computer. In the latter case the Hamiltonian dynamics are realized via Trotterization.

The basic approach discussed in this paper should be useful in situations in which it is comparatively easy to explicitly construct an initial state of a system that has a much higher energy density than the thermal system one wishes to study. In the context of the thermal properties of non-Abelian gauge theories, which serve as the key motivation of this paper, gauge invariance plays a central role. Thus, in describing the approach, we will emphasize places in which the preservation of gauge invariance is central and explain how gauge invariance is maintained. For applications to theories other than gauge theories this can be ignored; the rest of the algorithm goes through unchanged.

The implementation of a gauge theory in a Hamiltonian formalism is somewhat subtle compared to non-gauge theories. As one goes from the Lagrangian to Hamiltonian formulations, one faces the problem that in the Lagrangian formulation, the temporal component of the gauge field has no time-derivatives, meaning that it has no associated canonical momentum. Thus, one needs to choose a gauge-fixing condition to write a Hamiltonian. Perhaps the most natural of these for the purpose of quantum computing is the temporal gauge of $A_0=0$. Moreover, even after the gauge fixing is fixed, there is a residual gauge freedom that respects the gauge fixing condition.  

Note that the Euler-Lagrange equation obtained from the Lagrangian in this temporal gauge is not strictly an equation of motion as it does not involve time; rather, it yields the color version of the Gauss law and serves as an equation of constraint. States that are physical must satisfy the Gauss-law constraint. Thus the full Hilbert space of the theory is larger than the physical space. The Gauss-law constraint operator acts as a generator for residual gauge transformations which implies that all physical states of the theory are invariant under residual gauge transformations\cite{KogutSusskind}.

When it comes to the quantum simulation of gauge theories, it is important that any algorithm used has as an input that is a physical state and in its time evolution keeps the system in the physical space. This is nontrivial, as common digitization schemes \cite{digphi,diggluon} can create non-physical states. One way to assure physical states is to turn to gauge invariance, as physical states are exactly those states that are invariant under residual gauge transformations.  Fortunately, time-evolution commutes with gauge transformations, even under a Suzuki-Trotter approximation \cite{digsim,halimeh,symprot}. Thus, any state preparation scheme that begins with a gauge-invariant state and proceeds through time evolution will maintain a physical state, and thus respect gauge symmetry (up to noise, whose effects in pushing the system out of the physical subspace can be mitigated \cite{NuQS}).

As mentioned above, the motivation for this paper is specifically the study of various aspects of heavy-ion collisions, which face severe limitations in the world of classical computational methods. Thus, it is useful to ask what would be needed to obtain information crucial to the study of heavy-ion collisions if a sufficiently powerful quantum computer were available. The standard picture for heavy-ion collisions \cite{HIC1,HIC2} comes in three stages, each with their own theoretical description \cite{stages}. The first stage is the initial scattering of two high-energy nuclei, with collisions taking place at the partonic level. After this, a quark-gluon plasma (QGP) is formed, expands and cools, constituting the second stage. In the final stage, as the QGP cools sufficiently, the quarks and gluons hadronize. Notably, hydrodynamics has shown to be quite effective at modeling the QGP stage of the collisions \cite{hydro1,hydro2}. Data from LHC and RHIC \cite{LHC} can help model hydrodynamics for QGP, but there are many free parameters and uncontrolled systematics in the phenomenological models used to estimate them from the data; this adds uncertainty to analyses\cite{unfeasible1,unfeasible2}.

Of particular interest to the hydrodynamic study of heavy-ion collisions are the transport coefficients \cite{transpo1,transpo2,transpo3,transpo4,transpo5,transpo6}: diffusivity, conductivity, and viscosity. Previous research \cite{A} has proposed quantum algorithms to determine these transport coefficients, including working out the requisite Hamiltonian lattice operators. To evaluate the expectation values of these operators, two things are needed: a way to time-evolve these operators under the physical Hamiltonian, and states for the operators to act on.  A Trotterized time-evolution operator that respects gauge symmetries has been constructed in \cite{digsim}.

For transport coefficients, we want states to be chosen from a distribution corresponding to a system in thermal equilibrium. The aim of this paper is to propose an algorithm that can take a state that is straightforward to write on a quantum computer and show how to perform operations on it that allow for the sampling of energy eigenstates with approximately Boltzmann weights. In this way, thermal expectation values can then be estimated. 

One concern about thermal state preparation is the time-scale. There exist certain systems for which one expects thermalization to take a long time. An example: certain frustrated spin systems \cite{frustrated} at low temperatures are known to require state preparation times to be exponential in the volume. Fortunately, in the context of the principal concern here---heavy ion reactions in their QGP phase--- phenomenology strongly suggests that rapid equilibration occurs after the initial hard interactions \cite{QCD_crit_pt,HICattractors}. This equilibration is believed to occur on a time scale comparable to the QCD characteristic scale \cite{QCDscale}, and it is quite plausible that this quick equilibration is a general feature of strongly-coupled gauge theories. With this in mind, we expect developing a non-thermal state and letting it time-evolve will achieve approximate thermalization on approximately the natural time-scale of the theory. While this is far from a formal proof, this phenomenological argument makes it reasonable to assume that polynomial-time thermalization schemes are possible. Thus, we can proceed to the practical task of achieving such a scheme.

\subsection{State preparation methods}

There are many state preparation methods. It is believed that often the most arduous task in quantum simulation is this exact problem of preparing an initial state \cite{TSP1,TSP2,TSP3,TSP4,TSP5,TSP6,TSP7,TSP8}; thus, no ``one size fits all" scheme currently exists. Since quantum computing is naturally implemented via Hamiltonians, the goal of many state preparation methods is to prepare the system to be in the ground state of a chosen Hamiltonian.

Perhaps the most conceptually straightforward approach is a heat bath algorithm \cite{LargeHeatBath1}. It aims to construct a large heat bath in thermal equilibrium (at a finite temperature) and then couple the system to this heat bath weakly and allow energy to transfer until the system itself has equilibrated. The goal of this method is to create a state whose expectation value of a given observable reproduces the thermal expectation value of the observable at a certain temperature. This approach suffers from a large auxiliary system and long interaction times. The algorithm proposed in this paper aims to improve upon this basic approach. 

A classic scheme of ground state preparation is adiabatic state preparation \cite{fast_asp,gauge_asp,NMR_asp}, which aims to prepare the ground state of a desired Hamiltonian. This method begins with a Hamiltonian whose ground state is known and adiabatically time evolves the initial Hamiltonian into the Hamiltonian of interest. The adiabatic theorem \cite{messiah} tells us that this would obtain the desired ground state. It should be noted that when working with a gauge theory, care must be taken that gauge invariance is maintained in any approach. For field theories, it is natural to use lattice versions of the theory. One challenge faced by lattice gauge theories in a Hamiltonian formulation is the one alluded to earlier: that the Hilbert space of the theory is much larger than the physical Hilbert space\cite{KogutSusskind}. States in the physical space are constrained to satisfy the color Gauss law, which in a lattice formulation means that the net color electric flux flowing out of a lattice site must be equal to any color electric charge on the site. Equivalently, the state must be invariant under the residual gauge transformations permitted once a gauge-fixing condition that relates the temporal component of the gluon field (which has no conjugate momenta) to the spatial components is specified. In the adiabatic approach, if one can prepare a state that is the ground state of the physical Hilbert space of some limit of the theory and then adiabatically alter the theory while preserving gauge invariance, one is guaranteed that the system will remain in the physical Hilbert space of the system. For such gauge theories, an obvious starting state is the strong-coupling limit of the lattice gauge theory. The virtue of this limit is that an explicit prescription exists for the ground state in this limit in the physical space \cite{scott}. A concern is that noise or other sources of error may push the system out of the physical space; fortunately, as mentioned above, there is a technique that helps ameliorate this problem\cite{NuQS}. However, despite this technique, the long time evolution required for adiabatic state preparation is a major drawback:  quantum computers must constantly fight decoherence \cite{decohere}, and long times imply large opportunities for decoherence.

Another popular method of state preparation is via quantum phase estimation \cite{phaseest1,phaseest2,phaseest3}. This method starts with an approximate eigenstate and a unitary operator (often the time evolution operator). Then, through use of additional qubits (i.e. qubits not encoding the initial eigenstate) and operations of the unitary operator, the eigenvalue and true eigenvector are found to polynomial accuracy. This algorithm, however, requires that the time evolution be fully coherent. Variational methods \cite{varmeth1,varmeth2,varmeth3} of phase estimation have been developed to replace long time evolutions with a number of smaller time evolutions. However, these methods require an initial state with significant overlap to reconstruct the desired eigenstate, which means additional steps must be taken to get a good initial state which in turn can be used to construct the ground state. Moreover, finding an initial state with substantial overlap with the true ground state may well be exponentially hard as the size of the system increases.

A recently developed approach to state preparation is projective cooling \cite{projcool1,projcool2}. This method is similar in spirit to evaporative cooling. It acts to drive excitations out of the region of interest and into a larger complementary system. Then the wavefunction remaining in the region of interest is measured. A main drawback here is that the required complementary system must be significantly larger than the system of interest. A variation on this method is the Rodeo algorithm \cite{rodeo}, which uses ancilla qubit measurements to shed excitations, thus driving the initial state to a desired eigenstate. These algorithms depend on the overlap of the initial state and the desired eigenstate, making initial state choice quite important for efficiency. Again, a major drawback is that it becomes exponentially hard to find an initial state with substantial overlap with the true ground state as the size of the system increases.

 Resonance transitions \cite{resonance}, provide an alternative approach to state preparation. In this approach an ancilla qubit is coupled to the physical system. Then a resonance transition between the ancilla qubit and the physical system's ground state is driven and measurements on the ancilla qubit are performed to drive the physical system towards its ground state. This algorithm requires knowledge of the physical system's ground state energy, which poses a significant problem for lattice gauge theories.

 Finally, there are algorithmic cooling approaches. Heat bath algorithmic cooling \cite{heatbath2} pushes entropy onto a number of qubits that are then replaced by qubits that thermalize quickly with the environment; here, the environment is acting as a heat bath. This scheme was developed in the world of NMR quantum computing and has a fundamental limit of the temperature of the environment. Furthermore, it is not a universal method when it comes to cooling quantum many-body systems. Demon-like algorithmic cooling \cite{heatbath1} builds on this approach by using ancilla qubits to entangle with the system in such a way that measuring the qubit down (up) results in a system state of lower (higher) energy. While this is very similar to our approach in that energy is removed from the system via measurement, our algorithm has an important feature wherein the energy/temperature can be controlled.
 
 The proceeding approaches are aimed at finding the ground state. For thermal expectation values other state preparation methods are required.
 
 One approach is to use minimally entangled typical thermal states, or METTS \cite{METTS1,METTS2,METTS3}. This approach calculates thermal expectation values by starting with a basis state, evolving it in imaginary time, and then calculating the desired observables. Then this state is collapsed via measurement to a new basis state and the process is repeated. For reference, the state created after the imaginary time evolution is called a METTS. This approach must contend with correlations between samples; it also requires a large number of measurements to get a reasonable thermal average.


In this paper we develop another algorithm for the calculation of thermal expectation values. It is most closely related to ref.~(\cite{heatbath1}) in that it uses a ``Maxwell demon" type approach. Our approach has a number of virtues: It is designed to produce a thermal ensemble, and it can be implemented in a way that allows one to control the desired temperature.  It is also designed to be implementable on gauge theories without damaging gauge invariance. Moreover, this scheme is designed to have favorable scaling behavior when implemented on lattice theories with local interactions; the time taken to reach equilibrium at some temperature should be independent of the volume, while the total number of ancillary qubits needed grows linearly with the size of the system. Finally, this scheme might turn out to be useful in finding ground states, either via directly driving the temperature to zero or as a means to obtain sufficiently low temperatures so that states from the ensemble have a sufficiently large overlap with the ground state so that methods such as Rodeo algorithm \cite{rodeo} can be used efficiently to obtain the ground state.

This paper is organized as follows: Section II details our algorithm to construct Boltzmann distributions based on a method of active cooling. Furthermore, we argue that the energy of the system of interest can be predicted so that, given a mapping between energy and temperature, the Boltzmann distribution that is constructed will come from a chosen temperature range. This algorithm has been developed via the initial inspiration of ref.~\cite{A}. Section III describes a ``toy'' model that we used to illustrate this algorithm. Section IV discusses how to generalize our algorithm developed for the toy model to larger systems with a particular emphasis on large lattice models. We conclude in Section V with a discussion of the required future work.

\section{The Active Cooling Scheme}
The approach in this paper is an extension of the method outlined in ref.~\cite{A}. That scheme was designed to exploit the basic thermodynamic notion of a heat bath \cite{Thermo}, but to do so in a manner that does not require an overly large number of qubits. In a closed system with a heat bath approach, one uses a total system that is much larger than the subspace representing the system of interest. One would denote the system complementary to the system of interest as a heat bath. We assume that the subspace representing the physical system of interest is governed by Hamiltonian $H^{sys}$, and the heat bath is governed by a Hamiltonian $H^{HB}$ (which for gauge theories is defined according to a gauge-fixed scheme), which are initially decoupled. A key feature of $H^{HB}$ is that it is simple enough that an explicit form of its ground state is known and can be implemented easily on a quantum computer. The heat bath is initialized to its ground state and the system subspace is initialized in a physical state of higher energy than of interest---as it would be if it were in an easily produced physical state for a non-Abelian gauge theory.

The key point is that the specifics of the initial physical state of the system of interest should not matter provided that the heat bath is sufficiently large. Accordingly, we will simply use a state that is relatively simple to write on a quantum computer. For physical states in the system of interest, this will require a state with very high energy density if one is studying a non-Abelian gauge theory.

The subsystem for the system of interest and the heat bath would then be coupled via a Hamiltonian $H^{coup}$. $H^{coup}$ is derived from a Lagrangian that is gauge-invariant in the system of interest and is defined with the same gauge-fixing condition as $H^{sys}$, and thus it is invariant with respect to the same residual gauge transformations in the system subspace. Initially this coupling Hamiltonian is switched off and the system of interest is set in a known physical state, specifically one that satisfies the Gauss law. Since $H^{coup}$ is gauge invariant, Hamiltonian dynamics will ensure that the system will continue to obey the Gauss law and hence remain physical. 

Importantly, we require
\begin{align}
    [H^{coup},H^{sys}] \neq 0 \neq [H^{coup},H^{HB}]
\end{align}
This condition assures that energy will be allowed to transfer from the system subspace to the heat bath when the coupling is turned on. The next step is to switch on the coupling and allow the dynamics to run. To good approximation, the full system will eventually equilibrate with a temperature corresponding very closely to the one associated with the initial energy density of the heat bath.

The major defect of this heat bath approach is that the heat bath must be quite large for it to be effective. The logic of the heat bath is that it can thermalize with the system of interest without noticeably changing its own temperature.

Before describing alternatives to the large heat bath, it is useful to note that there are many challenges facing practical quantum computing for gauge theories.  The optimal method of calculating thermal expectation values may well depend on what is the most significant source of limitation for the problem under study; different approaches face different limitations. If the principle source of limitation is the number of well-connected qubits in the machine, then heat bath methods become problematic. The active cooling approach described below is a sensible way of dealing with this problem, but this may involve a trade off: it is possible that it will take longer to run than a single heat bath and that could prove problematic if the principal limitation is coherence time. 

With this consideration in mind, we continue examining our principal motivation in ref. \cite{A}. To understand the motivation for the scheme, one might first choose to imagine replacing a single large heat bath with many smaller disconnected ones, each initialized to their ground state. Acting together and left for a sufficiently long time, they would still act to thermalize the system subspace. There is no real gain in doing this as opposed to using a single large heat bath;  however, suppose that instead of connecting all of these heat baths simultaneously, one connected them sequentially and allowed each to act long enough to thermalize approximately before disconnecting it. Again one would expect that if one had a sufficient number of these small heat baths, the system would again thermalize to the desired temperature.

At this point one can ask whether instead of using many different heat baths, one can simply use the same heat bath over and over again. To do so, after disconnecting the small heat bath from the system of interest one would have to reinitialize it to its ground state after each use. The act of reinitializing the heat bath is non-unitary, since all states of the heat bath get mapped into its ground state. However, provided the ground state of the heat bath system is sufficiently simple that it can be explicitly specified, it is possible to do such a reinitialization; in fact, the DiVincenzo criteria \cite{DV} for the viability of a quantum computer requires this ability to reinitialize qubits to a specific state consistently. This state is denoted a fiducial state and here we will take the fiducial state of the heat bath to be its ground state. 

It is important to note that due to the cyclic nature of this scheme, we do not require that all the energy that must be removed from the system of interest be removed in one go; we only need that each cycle removes an appreciable amount of energy. Thus, we can use a heat bath that is much smaller than the large heat bath discussed above.

This process of repeatedly connecting and disconnecting the system of interest to a small heat bath, and reinitializing the heat bath to its ground state in between, is a form of active cooling. Rather than having energy simply passively flowing to a large heat bath, the step of reinitializing in effect transfers the extracted energy to the environment. Thus, the small heat bath is functioning as a refrigerator, and from here on it will be denoted as such and its Hamiltonian will be denoted $H^{Refrig}$.

\begin{figure*}[t]
\centering
\includegraphics[width=.68\textwidth]{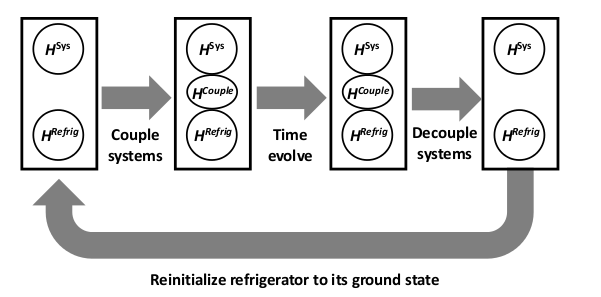}
    \caption{The active cooling cycle}  \label{fig:Active}
\end{figure*}

The active cooling proposed in ref.~\cite{A} is a cycle that contains four steps; the cycle is illustrated in Fig.~\ref{fig:Active}. In the first step, the refrigerator is initialized to its ground state and is not coupled to the system of interest. In the next step, the two systems are coupled via $H^{coup}$.  In the the third step, the coupled system is allowed to time evolve under the dynamics of $H^{sys}+H^{Refrig}+H^{coup}$ for some time. In a quantum simulator this would be done via direct Hamiltonian time evolution and in a gated quantum computer approximated via Trotterization.  Next the coupling Hamiltonian is switched off. Finally, the refrigerator system is reinitialized to its ground state and the cycle begins again.

If the coupling Hamiltonian is sufficiently weak that the energy contained in it is always negligible, then the cycle will always act to remove energy from the system of interest. This is because the refrigerator system starts in its ground state so its energy can only increase during the time evolution; on the other hand, energy is conserved in Hamiltonian time evolution so energy must flow from the system of interest to the refrigerator. The active step in the cycle is the act of reinitializing the refrigerator; it is in this step that the energy extracted from the system of interest is removed from the combined system. Note that if the goal is to remove energy from the system, the precise length of time that the system is coupled to the refrigerator is unimportant, as energy will always flow out of the system during a cycle.

Note that even if one were to start the first cycle with both the system of interest and the refrigerator in pure states, after a single cycle the combined system will be in a mixed state. The time evolution of the coupled system will induce entanglement between the system of interest and the refrigerator. It is instructive to understand how the entanglement can affect the system of interest, especially during the reinitialization step. The reinitialization can be done by measuring the refrigerator's energy and then conducting unitary transformations that return the refrigerator to its ground state. This measurement will collapse all the entanglements, which affects the reduced density matrix of the system of interest. The unitary transformations that follow, however, do not affect the physical system. 


If one's goal is to push the system of interest into a reasonable approximation of the ground state, this scheme should be a viable alternative to other state preparation methods and, in the future, could be investigated as such. In such an application, one would need to run a sufficient number of cycles to ensure the system was close enough to the ground state for the application under consideration. The motivation for proposing this scheme, however, was not to prepare the system in the ground state but rather for the eventual study of gauge theories at nonzero temperatures. In ref.~\cite{A}, it was suggested that this active cooling scheme could be used for such studies provided that they were done in the context of the microcanonical distribution.
 
The basic idea is that one could run through enough cycles to get the characteristic energy of the system in approximately the correct energy range.  The system would be in a mixed state, but its density matrix would not be an approximate thermal distribution. We see this since, even if the system and refrigerator were coupled together long enough to thermalize with each other, the reinitialization step would throw the system of interest out of thermalization due to entanglements. Furthermore, as mentioned above, one cycle is only tasked with removing an appreciable amount of energy. It seems intuitively clear that the time needed to remove a reasonable amount of energy is significantly smaller than the time needed to allow for thermalization. 

Thus, this algorithm is not aimed at constructing an approximately thermal state. It is, however, designed to construct a Boltzmann distribution on the ensemble level. That is to say that the state produced by the algorithm is such that measuring the energy will choose an energy eigenstate with a probability given by a Boltzmann weight. This comes about because the measurement inherent to the reinitialization step affects the system, as mentioned above; however, on average this change to the system is null. Thus, on the ensemble level the system is not pushed out of thermalization. In this way the algorithm constructs a Boltzmann distribution. 
 
This still leaves the matter of how to use this algorithm's thermal construction to, say, calculate thermal expectation values. This is straightforward, as one could directly measure the energy, casting the system into an energy eigenstate at known energy, and then measure the quantity of interest on that particular energy eigenstate. Such a procedure could be repeated and the results binned by energy with each bin covering a narrow range of energies. The quantity of interest could then be averaged over each bin. Since each bin corresponds to a narrow band of energies, this procedure would yield an approximate extraction of the microcanonical value for the quantity of interest. If the system is sufficiently large to well approximate the thermodynamic limit \cite{Thermo}, the value of the quantity at fixed energy in a microcanonical description will closely approximate its value in a canonical description at the temperature whose mean energy equals the microcanonical energy. 
 
Since we wish to compute observables as a function of temperature, a key step would be relate average energies to temperature. In principle, this can be done on a classical computer using standard Monte Carlo methods. One would fix $T$ and calculate the average $E$, repeating this process for a large number of temperatures. One could then invert this relationship to obtain a mapping from $E$ to $T$. 

Such a scheme may be rather efficient at removing energy. It was argued in ref.~\cite{A} that if  $H^{Refrig}$ and $H^{coup}$ were allowed to vary with the cycle and are optimally chosen, then, up to fluctuations, one ought to be able to achieve a fixed fractional energy loss per cycle. Thus the number of cycles would be
\begin{align}
    N^{cycles} &= A \log(\frac{E_i}{E_f})
\end{align} 
where $A$ is a constant and $E_i$ and $E_F$ are the initial and final energies. Thus, one could achieve very large fractional cooling in comparatively few cycles. 

As noted above, this approach is based on a kind of quantum refrigerator \cite{engine}. For a review of quantum refrigerators, see \cite{refrig}. Note also that we can call this a sort of quantum demon refrigerator, as it acts akin to Maxwell's demon \cite{demon1,demon2}. This is because of the reinitialization step of the refrigerator: this is obviously not a unitary operation and thus requires coupling to the environment. 

Such an approach should work; however, it has a few drawbacks.  One of these is that it is not straightforward to have the process yield states with an energy corresponding to a temperature in the range that one wishes to study. In practice, this may mean that in order to study anything, one must study a broad range of energies/temperatures rather than focus on getting good statistics on a handful of temperatures of interest. This could be problematic if the quantum computing resources are limited.
 
If the size of the system being studied is sufficiently large this may not matter, assuming that the eigenstate thermalization hypothesis (ETH)\cite{ETH1,ETH2,ETH3} is valid for the quantity of interest. the ETH implies that a single energy eigenstate for a large system should typically reproduce the thermal expectation value for a system whose temperature yields an average energy corresponding to the energy of the state. If the system were large enough for the ETH to hold, one would not need to acquire many states to do a statistical average, greatly alleviating the problem.  However, it is unclear whether the ETH will hold for quantities such as transport coefficients of field theories. Even if the ETH holds in principle, it is also unclear how large the system will need to be for it to be valid in practice. This latter issue is significant since the earliest practical calculations on quantum computers are likely to be on relatively small systems.

With this in mind, it would be greatly beneficial to have the ability to predict and control the characteristic energy of the system so that one could know when the system is in the energy range that is desired. This ability would allow for this active cooling scheme to be used efficiently to produce data in the desired range without many data points falling outside of said range.  

This is where a key benefit of the present scheme lies. Active cooling mainly consists of two steps: time evolution while the refrigerator and system of interest are coupled (which notably conserves energy) and the reinitialization step (which notably does not conserve the energy of the total system). Thus, if one keeps track of the energy that is removed during the reinitialization step, then the characteristic energy of the system of interest can be approximately known. This ability to track energy, plus the mapping between energy and temperature mentioned above, allows the construction of a Boltzmann distribution of a temperature lying in a chosen range. 

Keeping track of the energy removed is fairly straightforward: we assume, \textit{a priori}, that the ground state energy of the refrigerator is known. During the reinitialization step, one can measure the energy of the refrigerator and then transform it to its ground state. Given that the energy of the refrigerator before and after is known, the energy that was removed is also known. Note that the ability to do these procedures define the kinds of refrigerators that must be used for this active cooling scheme. A simple example of this is a refrigerator consisting of spins.

\subsection{Parameters}
The main parameters of this scheme are the size of the refrigerator, the parameters controlling the Hamiltonian of the refrigerator, the overall coupling strength (which we will denote $\sigma$), the number of cycles $N$, and the time that the refrigerator and the system of interest are coupled during each cycle, denoted $\tau$. It is straightforward to see that the number of cycles $N$ will be determined by the amount of energy that has been removed (which is monitored) and the target energy.

Here, we restate that due to the cyclic nature of this scheme, we do not require that the all the energy that must be removed from the system of interest be removed in one go; we only need that each cycle removes an appreciable amount of energy. Thus, we can use a refrigerator that is much smaller than the large heat bath discussed above. In fact, we claim that the refrigerator's size can be on the same scale as the system of interest, or even smaller, and still work efficiently. Again we will mention that this is subject to the trade off of qubit resources and coherence times. 

The studies with a ``toy model" in this work demonstrate that in fact a smaller refrigerator works well. Future work aims to show this for larger and more physically relevant systems. Future work also aims to study the optimal value of $\tau$, since the small systems studied in this paper have energy transfer occurring almost instantaneously. 

When it comes to the strength of the coupling, ideally we would want to impose a condition that $H^{coup}$ is small; specifically, we want a coupling strength $\sigma$ that is small enough so the coupling cannot appreciably give energy to the system subspace, as discussed above. This condition can be made explicit: we say $H^{coup}$ is sufficiently small when $\expval{H^{coup}} \ll \expval{H^{sys}}-\expval{H^{sys}}_{vac}$ holds at all times. Thus, an ideal choice of coupling strength would be the largest value of $\sigma$ that satisfies the above constraint. We want $\sigma$ to be as large as possible because stronger coupling allows for faster energy transfer.

It is important to note, at this point, that our ultimate goal is to develop a scheme that is blind to the system's energy spectrum since \textit{a priori} this is unknown. This means that we want to assume no knowledge of or ability to determine $\expval{H^{sys}}_{vac}$. Thus, we must dispense with this condition, at least \textit{a priori}, and deal with a coupling that has the ability to heat up the physical subsystem. In practice, $\sigma$ is chosen based off of a best guess of this ideal value.

A similar best guess approach is taken for the parameters that describe the refrigerator's Hamiltonian. Ideally, we want the average energy density of the levels of the refrigerator to match closely the average energy level density of the system of interest to promote efficient energy transfer. Again, we do not assume knowledge of this level density at all. 

At this point the selection of the refrigerator parameters as well as $\sigma$ seems to be almost blind, which is concerning as they are important parameters to the efficiency of this scheme. However, it is important to recall that this scheme starts with an initial state that has an energy much higher than the target energy since those states are the only ones we know how to straightforwardly write on quantum computers. Thus, we suggest that choosing a $\sigma$ that is not too big is not a difficult task. We also believe that picking refrigerator parameters that are serviceable is a similarly manageable task. Again, we point out that we do not require huge amounts of energy removed every cycle; we simply want appreciable amounts of energy removed. Said another way, we claim that this algorithm is relatively robust in that it is not overly sensitive to the initial parameter inputs.

In this paper we develop a detailed algorithm in the context of a simple ``toy" model to illustrate how such a scheme can work. Before describing this algorithm, we want to highlight the beneficial aspects of this algorithm. First and foremost, it is capable of tracking the energy of the system to reasonably accurately aim at the desired characteristic energy and thus, with the $E$ to $T$ map mentioned above, the desired temperature. Also, unlike the projective cooling algorithms, its complementary system can be of a size on the same order of, instead of significantly larger than, the size of the system subspace. We also utilize the complementary system iteratively to significantly reduce the amount of time required. Moreover, the algorithm was designed to work for any initial state and without any knowledge of the energy spectrum. The algorithm has two other useful features that apply to its use on larger systems. One is that the refrigerator can be used a thermometer to both test whether the system has achieved thermal equilibrium and if so at what temperature. The second is that the system can thermalize (to good approximation) at an early stage and should remain close to equilibrated throughout. These aspects are discussed in more detail in Section IV. 

While this algorithm was developed and illustrated in the context of a particular toy model, it is highly plausible that it should work quite generally. Moreover, as will be discussed later, there are good reasons to believe that the model should have good scaling properties when applied to lattice models with local interactions.
 
\section{A Toy Model}

In testing our algorithm, we want to demonstrate its applicability to different Hamiltonians, so we want to test it on a class of Hamiltonians instead just one. 
Thus, we will not present a singular Hamiltonian for our ``toy" model but rather the method we used to generate it. We claim that our algorithm works for every Hamiltonian this method generates. The results portrayed in subsection B pertain to only one of these Hamiltonians, but similar results can be obtained from other Hamiltonians generated from the method described below.

In constructing the Hamiltonian for our ``toy" model, we wanted to pick a rather generic Hamiltonian but we also wanted to mimic the property of physical systems wherein the level density increases exponentially with energy. Thus, we handcrafted an energy spectrum that has the latter property by picking an energy span (0 to 1000 energy units), identifying 10 regions of 100 energy units each, and using the function $round(e^{0.8\sqrt{i}})$ (where $i$ ranged from 0 to 9) to determine the number of energy levels in each region; these levels were evenly spaced in their respective energy regions. With this energy spectrum in hand, we moved to a randomly chosen eigenbasis for coupling reasons that will become clear below. Finally, to introduce some randomness to the system's Hamiltonian, we added to our existing Hamiltonian a randomly generated matrix where each entry has chosen from a normal distribution with mean 0 and a standard deviation that is small enough to not destroy the exponentially increasing level density.

We note here that the system was chosen to have 57 levels, as that is enough levels to be able to appreciably see the desired exponential increase in level density while also not overtaxing the classical computers on which simulations of this system were conducted. We also note that since this algorithm was developed for QCD problems that obey time-reversal symmetry, we made sure that the system Hamiltonian was symmetric. 

\subsection{The algorithm}

For the refrigerator, we used a simple spin model with 4 spins uncoupled to each other:
\begin{equation}
    H^{Refrig} = \sum\limits_{i=1}^s h_i\sigma_{z,i}
\end{equation}
where $s=4$ is the number of spins in the refrigerator, $\sigma_{z,i}$ is the z Pauli matrix for the $i$th spin, and $h_i$ is the field strength for the $i$th spin  ({\it i.e.} it fixes the energy difference between up and down). Note that the $h_i$'s are chosen to be positive, so that the ground state is when all the spins are down. Note also that since we are using 4 spins, $H^{Refrig}$ is an 16x16 matrix as opposed to $H^{sys}$ which is a 57x57 matrix. This represents a refrigerator that is significantly smaller than the system of interest. Clearly for larger systems it would make sense to use more than four spins.

We used a different field strength for each spin to ensure that correlations between the energies of the spin do not distort the results. The phenomenon of concern is where if there are two levels in the combined system with very close energies (where one level arises from a higher energy level in the system and a lower energy level in the refrigerator while the other level arises from a lower energy level in the system and a higher energy level in the refrigerator), energy transfer from the system to the refrigerator will proceed preferentially by the combined system moving from one of these levels to the other. This effect favors certain energy levels in the system of interest, which we want to avoid to promote equilibration. Thus, we used a different field strength for each spin to suppress these effects. We suspect that, for larger systems, this effect will become negligible and thus a single field strength for all the spins might well be used. For similar reasons, and also to encourage more efficient energy transfer, we change these field strength values periodically throughout the algorithm's run. We will be more explicit about this below.

The reinitialization step of a cycle is conducted by picking a singular spin and measuring it, which forces it into either $\ket{\uparrow}$ or $\ket{\downarrow}$. If it was measured up, it is then flipped down via $\sigma_{x,i}$; if it was measured down, the reinitialization step is complete. Note that we are only measuring one spin per reinitialization step instead of reinitializing all spins. This is because flipping one spin is much less of a disturbance to the total system than flipping all of the spins. The idea is to flip one spin at a time and while flipping one spin, the physical subsystem can remain coupled to the remaining spins. If the system were to achieve thermal equilibrium, flipping a single spin would only minimally disturb it. Thus, if thermal equilibrium is reached, the system can subsequently be kept in quasi-thermal equilibrium as the algorithm proceeds. Our toy model is too small to demonstrate this benefit, but it is straight forward to see that a single spin flip will be a small perturbation for larger systems.

At this point we will make it clear that there are three main sources of energy change for this algorithm: 1) the energy lost from flipping a spin down; 2) the energy change due to measuring a spin (either up or down); and 3) the energy change due to changing the $h_i$'s. The first source of energy change is the most controlled: we set the parameters of the refrigerator and thus we know exactly how much energy is lost when flipping a spin from up to down (namely $2h_i$). We keep track of this energy loss so that the energy of the system can be approximately predicted. 

The second source of energy change, arising from spin measurement, is the main source of uncertainty when it comes to predicting the energy of the system of interest. Note, however, that the energy changes arising from many spin measurements should average to 0 (holding the field strengths constant). Thus, this source of error in the physical subsystem's energy prediction should be suppressed for larger systems, as they naturally require more spin measurements. 

The third source of energy change is due to manipulation of the refrigerator parameters. This energy change comes because when the field strength values $h_i$ are changed, the probability of the refrigerator being in any of its energy levels is not changed but of course the energy of the levels is changed. For this toy model, each time we chose the $h_i$'s we chose them from a uniform distribution with fixed endpoints. It makes intuitive sense that with more spins in the refrigerator, the changes in the energy expectation value of the refrigerator will be less severe. This source of energy change is another source of error in the prediction of the physical subsystem's energy, but for the reasons just stated we believe this error is also suppressed for larger systems, which come with larger refrigerators. 

Finally, we now define the coupling scheme for our toy model.  Unlike the choice of Hamiltonian for the refrigerator, the coupling Hamiltonian described here is specific to the toy model. As will be discussed below, the appropriate coupling Hamiltonian for lattice models is straightforward to construct: each spin in the refrigerator is coupled to a small region of space of the physical subsystem.
\begin{align}
    H^{coup} = \sum\limits_{i=1}^s \sigma_{x,i}\otimes K_i
\end{align}
For the toy model, the coupling Hamiltonian $H^{coup}$ is defined as in (4), where the $i$th spin in the refrigerator is coupled to the physical system via a Kronecker product of $\sigma_{x,i}$ in the refrigerator subspace and a matrix $K_i$ in the system subspace. Each matrix $K_i$ is defined as follows: a 57x57 matrix (same size as the system Hamiltonian) whose entries are all 0 except for a randomly chosen 3x3 block. This 3x3 block has entries chosen from a normal distribution with mean 0 and standard deviation $\sigma$ which defines the coupling strength. Also, to ensure time-reversal symmetry is again respected while keeping the Hamiltonian hermitian, the transpose of this matrix is added to itself before the Kronecker product in (4) is taken. Note that the coupling is via the x Pauli matrix; this is done so that, when the refrigerator is in its ground state (i.e. all spins down), the expectation value of the coupling Hamiltonian is zero. 


This 3x3 block coupling scheme was chosen to be as close to analogous as possible to the case of lattice models wherein each spin in the refrigerator is coupled to a small region of space. We use the random eigenbasis for the system Hamiltonian to make this 3x3 block coupling effective.

Now with all the parts of our toy model defined, we can delineate a run of the algorithm. The system is initialized to its highest energy state while the refrigerator is initialized to its ground state. Parameters $\sigma$ and $\tau$ are chosen. When choosing $\sigma$, we ran multiple values from a wide range to assure that the algorithm is not too sensitive to this initial input. As discussed above, we reset the field strengths $h_i$ multiple times. Each time, they are randomly generated from a uniform distribution whose endpoints are chosen at the outset. These endpoints are chosen so that there is a large enough gap to give a good range of $h_i$'s so that even if some are too small/large, some will be much more reasonable. On the flip side, a larger range of possible $h_i$'s will increase the error due to this change, so it should not be made too wide a range. These considerations determine the choices for these endpoints.

Now, between reinitializations, we reset $h_i$ as well as the coupling (meaning the positions of the 3x3 blocks and the entries of these blocks are regenerated randomly), $p$ times. For this toy model, we used $p = 4$. This process of resetting the refrigerator parameters and the coupling is done to promote efficient energy transfer. Thus, the system is coupled to the refrigerator for a time $\tau/p$, then they are decoupled and the refrigerator parameters and coupling are reset. This is repeated until a full time $\tau$ has passed with the system and the refrigerator coupled to each other. At this point, the reinitialization step occurs: a single spin is chosen, measured, and flipped down if it was measured up. The spins are chosen in cyclical manner: 1,2,3,4,1, etc. (where the spins are indexed between 1 and $s$).

This process is repeated $N$ times. The coupling strength $\sigma$ is set for each cycle as $\sigma(t) = \frac{A}{1+Bt^{\alpha}}$ where $A$ and $B$ are constants that determine the starting and ending $\sigma$s, $\alpha$ determines the speed of decrease,and $t$ ranges from 1 to $N$. For the last iteration, the reinitialization step is slightly different: since we want to compare the initial energy to the current energy of the system, we want the refrigerator to be in its ground state both times. Thus, we reset all $s$ spins instead of just one. Finally, the energy of the system is measured, forcing the system into a singular energy eigenstate. 

Before presenting results from numerical experiments, we note that the choices made in developing the algorithm for this toy model was aimed at illustrating the concept with small systems. As mentioned, future work aims to show that the concept of this algorithm can work efficiently for larger systems. 

\subsection{Results}
We ran our algorithm on an ensemble of 1250 copies of a system chosen in the way described above. Throughout each run, we kept track of the energy that was removed due to spin flips. At the end of each run, the energy of the system of interest was measured. Then we compared our energy estimate (initial system energy minus the energy removed via spin flips) with the energy measurements. As discussed above, there are other sources of energy change in the system besides the spin flips, but for larger systems these sources will be suppressed; thus, we will compare both our energy estimates defined above and the expectation value of the system with the energy measurements to show the algorithms ability to control the energy. We stress here that in reality, taking the expectation value of the system is quite costly, and we are in no way including it as part of our algorithm; we are simply using the expectation value of the system energy as a proxy for the energy estimate (initial energy minus spin flip energy) for a larger system which will have other energy changes suppressed. 

For reference, the system used has an energy span of about 1000 energy units. The average difference between the energy estimate and the energy measurements was $-60 \pm 269$ energy units; The average difference between the expectation value of the system and the energy measurements was $11 \pm 203$ energy units.

To determine if these standard deviations are consistent with what we expect, we must first turn to our claim of thermalization. Assuming the ensemble is thermalizing (an assumption that will be justified below), we can estimate the temperature by calculating an average energy from the energy expectation values. From this average energy and a knowledge of the energy levels, we can extract a temperature. We note here that this algorithm does not require knowledge of all energy levels, it only requires a mapping from energy to temperature as discussed above. This procedure provides a temperature of $218$ energy units. 

From this temperature, we then sample multiple energy eigenstates from a Boltzmann distribution of this temperature. The standard deviation of these sampled energies was 232 energy units. This puts our standard deviations above in perspective: our standard deviations are consistent with what one would expect when sampling from a Boltzmann distribution at a temperature of $218$ energy units. 

\begin{figure}[h]
    \centering
    \includegraphics[width=\linewidth]{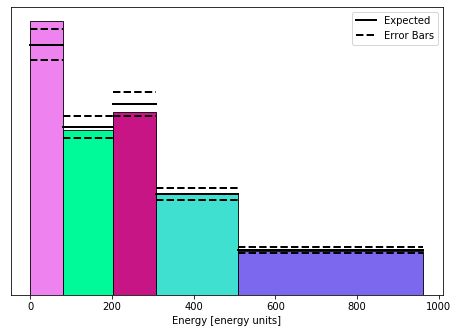}
    \caption{Data (colored bars) vs Expected (black lines)}
    \label{fig:Temp}
\end{figure}

Now that we have shown the algorithm's ability to monitor the energy, we turn to proving the assumption that we do in fact have a Boltzmann distribution. These results can be seen in Figure \ref{fig:Temp}. From our 1250 runs, we bin the measured energies into 5 bins. These bins were constructed so that a perfect Boltzmann distribution of a temperature $218$ energy units would have $20\%$ probability for each bin. The bins are designed to each have equal area; as such the heights are irrelevant and so the y-axis labels have been left off Figure \ref{fig:Temp}. 

As Figure \ref{fig:Temp} shows, the data (represented by the colored bars) lie within 1 standard deviation in four out of the five bins; the fifth bin lies within 2 standard deviations. Thus, we conclude that we have in fact constructed a Boltzmann distribution with our algorithm.

\section{Generalization to larger systems}

We have developed this algorithm in the context of a small simple toy model. It is hoped that quantum computers will continue to develop to the point that they can be used for much larger systems of physical interest. The algorithm developed here should work for such systems. Additionally, in larger systems some refinements are possible that should act to make it both more accurate and more controlled.  Clearly one would want a refrigerator with more than four spins, but it would still be useful to have them uncoupled and measured and flipped as needed to lower the energy.

As noted above, it is plausible that one does not need to randomize the field strengths of the refrigerator for large systems as the correlations leading to spurious results become increasingly less likely. However, there is very little cost to randomizing them.

\subsection{The refrigerator as a thermometer}
For large systems, one expects the system to thermalize at least approximately relatively early in the process. Indeed, if the initial state is very high energy and refrigerator is relatively small it is likely to thermalize at a negative temperature. Thus, there is a potentially useful alternate version of the proposed algorithm wherein only one spin at a time is reinitialized instead of the entire refrigerator. Since each spin flip only removes a very small fraction of the total energy, one expects that shortly after a spin flip, the system remains in approximate equilibrium with nearly the same temperature. Thus, by looking at the probability that a spin is measured up over a number of successive spin measurements, one should be able to determine an approximate temperature.

The obvious advantage to this is that in thermal studies of large systems we typically wish to study a system at a known fixed temperature as opposed to a fixed energy density.  Continuously monitoring the temperature is a very simple way to do this---one simply runs the algorithm until the system reaches the temperature of interest and then stops.

It is noteworthy that one needs to monitor the probability that spins are up or down in any case if one wants to run the algorithm efficiently. As discussed above, the field strength can neither be too big or too small for the efficient running of the algorithm.  If it is too large, the probability of a spin flip is very low and rapid energy transfer from the system to the refrigerator is inhibited.  Conversely, if it is too small, then spin flips can occur easily but each one deposits very little energy and again the energy transfer is inhibited. Since whether it is too big or too small (in this sense) varies with time as the algorithm runs, one needs to change the field strength as the algorithm evolves. A useful way to tell whether the field strength is in the ``Goldilocks'' region where it is neither too large nor too small is to focus on the temperature or equivalently the average probability of spin up. The Goldilocks region is typically  where the field strength is similar to $|T|$, where the absolute value sign accounts for negative temperatures. This criterion is useful except in the immediate vicinity of the region where $T$ diverges. As the system moves out of the Goldilocks region one needs to adjust the field strength.

\subsection{Lattice models}

Lattice models of formally infinite systems play an important role in both condensed matter physics and in nuclear and particle physics where they are used as proxies for continuous field theories by choosing the lattice spacing to be much smaller than the relevant physical scales in the problem.  Such models are studied via numerical simulation and they are studied on lattices whose physical extent is large compared to the relevant physical scales. Note that the present algorithm was principally developed to describe thermal properties of QCD and related gauge theories which are typically treated numerically as lattice models.

In many ways, the algorithm is better suited to lattice models than the toy model considered here since lattice models tend to thermalize locally. Suppose that one couples spins in the refrigerator to parts of the system in fixed spatial regions of the lattice model. If the model was in thermal equilibrium prior to a measurement of the refrigerator, the effect of a measurement of a single spin can at worst effect the equilibrium of the system in the region coupled to the spin and regions close to it. Thus, if one does not make another measurement of a spin that is coupled to a region nearby within a time characteristic of thermalization, then after the system equilibrates once, all future measurements will be of an equilibrated system.  

Accordingly, when implementing the algorithm on a lattice models one should couple the spins in the refrigerator to particular spatial regions of the model. Moreover, unlike in the toy model, there is no reason to randomly change the states or regions to which a given spin is assigned. Rather one should sample the the various regions in a manner that ensures that one does not sample a region near to one that has previously been sampled unless a sufficient time has elapsed to allow the system to equilibrate locally.  One can do this for example by randomly sampling spins coupled to a region subject to a constraint that no spin coupled to a nearby region has been sampled within a prescribed time.

\subsection{Algorithm for large lattice systems}

While it is too expensive to simulate large lattice systems on classical computers, it is instructive to outline how the algorithm would work for large lattice systems. The refrigerator will still consist of uncoupled spins; these spins will be coupled to localized regions of the system of interest. These localized regions are to be spaced out, ideally to "cover" the entirety of the physical system. There are many ways to define the coupling of the spins to the lattice system; of course, this must be done in a gauge invariant way. A simple example would be to couple spins in the refrigerator via $\sigma_x$ to a single spatial plaquette in the lattice system. In theory, any gauge invariant scheme in the lattice system could work, including any size of spatial Wilson loop, and empirical studies would need to be conducted to illuminate the most efficient form of coupling. 

With a refrigerator and coupling scheme in hand, the algorithm proceeds with a number of cycles. These cycles are to be slightly different from the ones diagrammed in Fig. \ref{fig:Active}: while the physical system and the refrigerator are still to be coupled and then the combined system time evolved, we do not require the full decoupling of the refrigerator with the physical system. Instead, we propose that one spin at a time is decoupled from the physical system and reinitialized (i.e. measured and if it was measured up, flipped down). We take this one-spin-at-a-time approach so that the physical system can, in the meantime, continue to evolve with all the other spins. The selection of the spin to be reinitialized should be done in the way discussed above: if a spin from a certain neighborhood is reinitialized, another spin from this neighborhood will not be reinitialized for an amount of time that should allow for the system to locally equilibrate again.

Furthermore, the results from the reinitialization steps should be recorded. The energy removed from the total combined system can be monitored, as is done with our small toy model, but also the results of spin up measurements should be recorded. As discussed above, this allows for the refrigerator to be used as a thermometer: the average probability of measuring a spin up relates to a temperature. With this tracking ability, one can stop the time evolution (by decoupling all the spins) once the target temperature is reached.

\section{Conclusion}
We have outlined an algorithm developed to work for a generic initial state and system of interest that produces a good way to sample energy eigenstates from a thermal distribution so that thermal expectation values can be calculated, and we have shown that is effective for small systems.  We have outlined how the approach should work for larger systems described by lattice field theories.  Future work aims primarily to test this algorithm for larger systems. With this, other variations of the algorithm can be tested out, including different refrigerator Hamiltonians. Future work also aims to test this algorithm on simple gauge theory models, to show that this algorithm may be effective for the specific problems it was motivated to address.

\begin{acknowledgments}
This work was supported in part by the U.S. Department of Energy, Office of Nuclear Physics under Award Number(s) DE-SC0021143 and DE-FG02- 93ER40762.
\end{acknowledgments}

\bibliography{biblio.bib}
\end{document}